\documentclass[sigconf]{acmart}
\usepackage[utf8]{inputenc}

\usepackage{caption}
\usepackage{algpseudocode}
\usepackage{amsmath}
\usepackage{amsfonts}
\usepackage{subfigure}
\usepackage{graphicx}

\usepackage{amsfonts}
\usepackage{booktabs}
\usepackage{siunitx}

\usepackage{arydshln}

\usepackage{color}

\usepackage{tikz}

\usepackage{enumitem}

\usepackage{url}
\makeatletter
\def\url@leostyle{%
  \@ifundefined{selectfont}{\def\UrlFont{\sf}}{\def\UrlFont{\small\ttfamily}}}
\makeatother
\newcommand{\eat}[1]{}
\usepackage[linesnumbered,ruled]{algorithm2e}
\usepackage{mdframed} 

\usepackage{xcolor}
\definecolor{light-gray}{gray}{0.9}

\usepackage{amsthm}

\newcolumntype{L}[1]{>{\raggedright\let\newline\\\arraybackslash\hspace{0pt}}m{#1}}

\title{SteemOps: Extracting and Analyzing Key Operations in Steemit Blockchain-based Social Media Platform}

\author{Chao Li}
\affiliation{%
  \institution{
  Beijing Key Laboratory of Security and Privacy in Intelligent Transportation \\
  Beijing Jiaotong University}
  \city{Beijing}
  \country{China}
}
\email{li.chao@bjtu.edu.cn}

\author{Balaji Palanisamy}
\affiliation{%
  \institution{School of Computing and Information \\University of Pittsburgh}
  \city{Pittsburgh}
  \country{USA}
}
\email{bpalan@pitt.edu}

\author{Runhua Xu}
\affiliation{%
  \institution{IBM Almaden Research Center}
  \city{San Jose}
  \country{USA}
}
\email{runhua@ibm.com}

\author{Jinlai Xu}
\affiliation{%
  \institution{School of Computing and Information \\University of Pittsburgh}
  \city{Pittsburgh}
  \country{USA}
}
\email{jinlai.xu@pitt.edu}

\author{Jingzhe Wang}
\affiliation{%
  \institution{School of Computing and Information \\University of Pittsburgh}
  \city{Pittsburgh}
  \country{USA}
}
\email{jiw148@pitt.edu}

\copyrightyear{2021}
\acmYear{2021}
\setcopyright{acmcopyright}\acmConference[CODASPY '21]{Proceedings of the Eleventh ACM Conference on Data and Application Security and Privacy}{April 26--28, 2021}{Virtual Event, USA}
\acmBooktitle{Proceedings of the Eleventh ACM Conference on Data and Application Security and Privacy (CODASPY '21), April 26--28, 2021, Virtual Event, USA}
\acmPrice{15.00}

\begin{document}

\begin{abstract}

Advancements in distributed ledger technologies are driving the rise of blockchain-based social media platforms such as \textit{Steemit}, where users interact with each other in similar ways as conventional social networks. These platforms are autonomously managed by users using decentralized consensus protocols in a cryptocurrency ecosystem.
The deep integration of social networks and blockchains in these platforms provides potential for numerous cross-domain research studies that are of interest to both the research communities.
However, it is challenging to process and analyze large volumes of raw \textit{Steemit} data as it requires specialized skills in both software engineering and blockchain systems and involves substantial efforts in extracting and filtering various types of operations.
To tackle this challenge, we collect over 38 million blocks generated in \textit{Steemit} during a 45 month time period from 2016/03 to 2019/11 and extract ten key types of operations performed by the users.
The results generate \textit{SteemOps}, a new dataset that organizes more than 900 million operations from \textit{Steemit} into three sub-datasets namely
(i) social-network operation dataset (SOD),
(ii) witness-election operation dataset (WOD) and
(iii) value-transfer operation dataset (VOD).
We describe the dataset schema and its usage in detail and outline possible future research studies using \textit{SteemOps}.
\textit{SteemOps} is designed to facilitate future research aimed at providing deeper insights on emerging blockchain-based social media platforms.

\end{abstract}



\maketitle

\section{Introduction}

Rapid development of distributed ledger~\cite{nakamoto2008Bitcoin} technologies is driving the rise of blockchain-based social media platforms, where users interact with each other in similar ways as conventional social networks. 
These platforms are autonomously managed by users using decentralized consensus protocols in a cryptocurrency ecosystem.
Examples of such platforms include \textit{Steemit}\footnote[1]{https://steemit.com/}, \textit{Indorse}\footnote[2]{https://indorse.io/}, \textit{Sapien}\footnote[3]{https://beta.sapien.network/} and \textit{SocialX}\footnote[4]{https://socialx.network/}.
Among all these platforms, \textit{Steemit} has kept its leading position since its launching in 2016/03 and its native cryptocurrency, \textit{STEEM}, has the highest market capitalization among all cryptocurrencies issued by blockchain-based social networking projects.
Today, \textit{Steemit} is considered as one of the most successful blockchain-based applications.

\textit{Steemit} enables the deep integration of social networks with the underlying blockchain infrastructure.
In \textit{Steemit}, users can perform various types of social-network operations as in Reddit~\cite{stoddard2015popularity} and Quora~\cite{wang2013wisdom}, such as creating blog posts, upvoting posts or comments and following other users.
Meanwhile, all data generated by \textit{Steemit} users are stored in its backend Steem-blockchain~\cite{Steem_blockchain} based on Delegated Proof of Stake (DPoS) consensus protocol~\cite{larimer2014delegated}. Users perform witness-election operations to periodically elect block producers called witnesses, as well as value-transfer transactions to transfer cryptocurrencies as in Bitcoin~\cite{nakamoto2008Bitcoin} and Ethereum~\cite{buterin2014next}.
Interestingly, different types of operations often correlate with each other. For instance, a user who aims at becoming a block producer may leverage the social network to advertise and promote himself or herself, as well as use cryptocurrencies to bribe important voters. As all  relevant operations are stored in the Steem-blockchain, it is available to the public and hard to be manipulated.
Therefore, the joint analysis of various types of operations in \textit{Steemit} provides potential for numerous cross-domain research studies that are of interest to both the social networking and blockchain research communities~\cite{thelwall2017can,kiayias2018puff,kwon2019impossibility,li2019incentivized,li2020comparison}.

Processing and analyzing large volumes of raw data in Steem-blockchain for creating useful datasets involves several challenges.
First, it requires sophisticated knowledge in understanding the Steem-blockchain, including but not limited to its DPoS consensus mechanism, cryptocurrency ecosystem and their associations with social behaviors in \textit{Steemit}.
As the white paper on \textit{Steemit} only provides limited information, it is necessary to consult a large number of technical articles posted by the development team, investigate the source code of the platform and register a few real accounts to match the frontend operations with the backend data in the Steem-blockchain.
Second, it involves substantial efforts in extracting and filtering various types of operations. 
The Steem-blockchain generates one block every three seconds and each block may contain over thirty different types of operations. 
While the billions of operations in the blockchain include a great deal of useless information, it is necessary though difficult to filter out undesirable operations.

To tackle this challenge, we collect over 38 million blocks generated in \textit{Steemit} during a 45 month time period from 2016/03 to 2019/11 and extract ten key types of operations performed by the users.
The results generate \textit{SteemOps}, a new dataset that organizes over 900 million operations from \textit{Steemit} into three sub-datasets:
1) social-network operation dataset (SOD);
2) witness-election operation dataset (WOD);
3) value-transfer operation dataset (VOD).
We describe the dataset schema and its usage in detail and outline various potential research directions based on \textit{SteemOps}.
\textit{SteemOps} is designed to facilitate future studies aimed at providing better insights on emerging blockchain-based social media platforms.


\section{Background}

In this section, we introduce the background about the Steem-blockchain~\cite{Steem_blockchain}, including its key application \textit{Steemit}, its implementation of the DPoS consensus protocol and its ecosystem in general.

\begin{figure}
\centering
{
    \includegraphics[width=8cm,height=6.5cm]{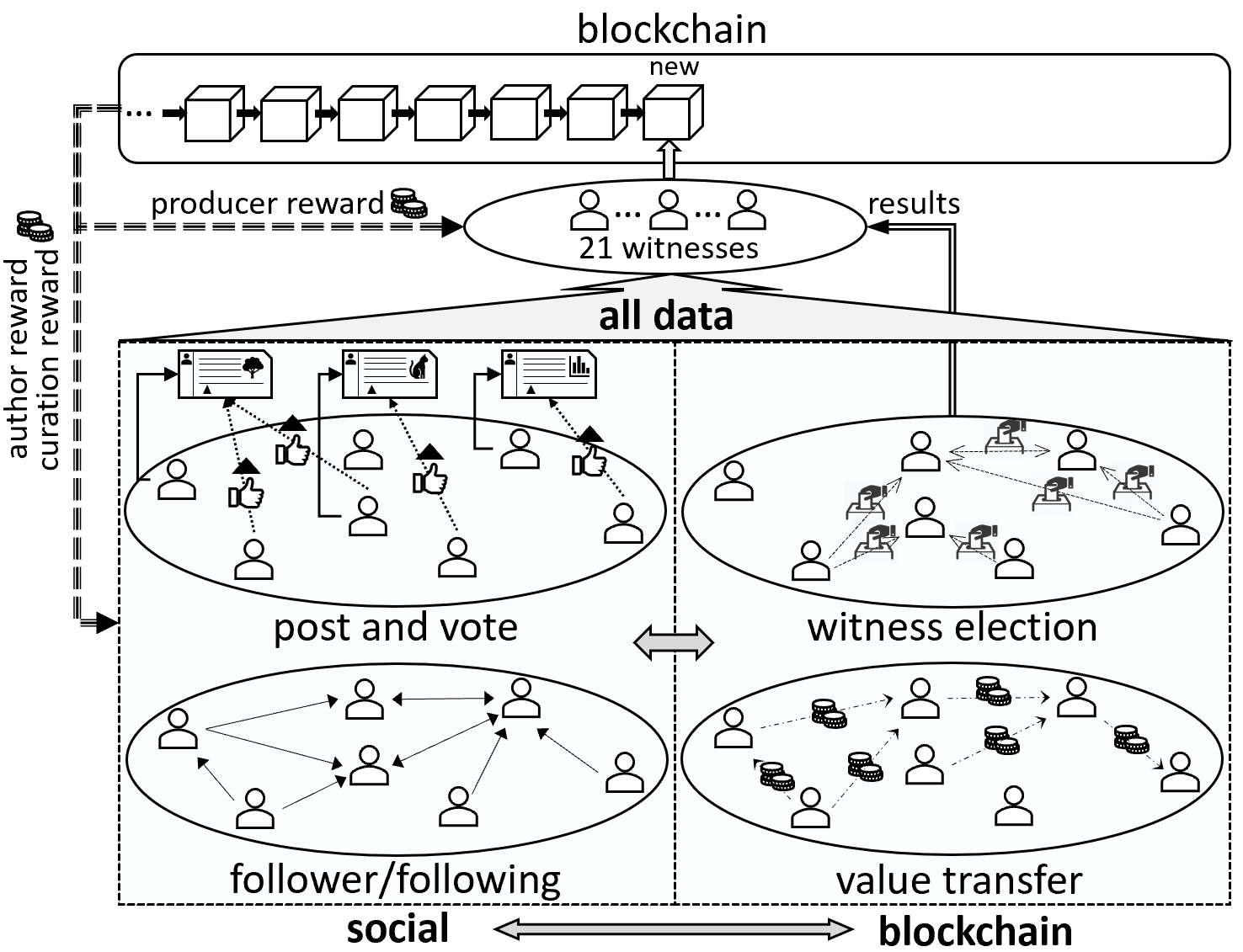}
}
\vspace{-1mm}
\caption {Steem blockchain overview}
\vspace{-6mm}
\label{f2} 
\end{figure}

In \textit{Steemit}, users can create and share contents as blog posts.
A blog post can get replied, reposted or voted by other users.
Based on the weights of received votes, posts get ranked and the top ranked posts make them to the front page. 
\textit{Steemit} uses the Steem-blockchain to store the underlying data of the platform as a chain of blocks. Every three seconds, a new block is produced, which includes all confirmed operations performed by users during the last three seconds.
\textit{Steemit} allows its users to perform more than thirty different types of operations. 
In Fig.~\ref{f2}, we display representative types of operations in \textit{Steemit}.
While post/vote and follower/following are common features offered by social sites, 
operations such as witness election and value transfer are features specific to blockchains.

Witnesses in \textit{Steemit} are producers of blocks, who continuously collect data from the entire network, bundle data into blocks and append the blocks to the Steem-blockchain.
The role of witnesses in \textit{Steemit} is similar to that of miners in Bitcoin. 
In Bitcoin, miners keep solving Proof-of-Work (PoW) problems and winners have the right to produce blocks.
However, with PoW, Bitcoin achieves a maximum throughput of 7 transactions/sec ~\cite{croman2016scaling}, which is too low for a social site.
Hence, the Steem blockchain adopts the Delegated Proof of Stake (DPoS)~\cite{larimer2014delegated} consensus protocol to increase the speed and scalability of the platform without compromising the decentralized reward system of the blockchain.
In DPoS systems, users vote to elect a number of witnesses as their delegates. In \textit{Steemit}, each user can vote for at most 30 witnesses. The top-20 elected witnesses and a seat randomly assigned out of the top-20 witnesses produce the blocks.
With DPoS, consensus only needs to be reached among the 21-member witness group rather than the entire blockchain network like Bitcoin, which significantly improves the system throughput.

The cryptocurrency ecosystem in Steem includes some complex features.
Like most blockchains, the Steem-blockchain issues its native cryptocurrencies called \textit{STEEM} and Steem Dollars (\textit{SBD}).
To own stake in \textit{Steemit}, a user needs to `lock' \textit{STEEM}/\textit{SBD} in \textit{Steemit} to receive Steem Power (\textit{SP}) at the rate of $1\ STEEM = 1\ SP$ and each $SP$ is assigned about 2000 vested shares (\textit{VESTS}) of \textit{Steemit}.
A user may withdraw invested \textit{STEEM}/\textit{SBD} at any time, but the claimed fund will be automatically split into thirteen equal portions to be withdrawn in the next thirteen subsequent weeks.

\begin{table}
\small
\begin{center}
\begin{tabular}{|p{2.7cm} |p{5.0cm}|}
\hline
{\textbf{OP (social-network)}} & {\textbf{Description}} 
\\ \hline
    comment & users create posts, reply to posts or replies \\
    vote & users vote for posts \\
    custom\_json & users follow other users, repost a blog \\ 
    \hline
{\textbf{OP (witness-election)}} & {\textbf{Description}} 
\\ \hline
    witness\_update & users join the witness pool to be elected, witnesses in pool update their information \\
    witness\_vote & users vote for witnesses by themselves \\
    witness\_proxy & users cast votes to the same witnesses voted by another user by setting that user as their election proxy \\
    \hline
{\textbf{OP (value-transfer)}} & {\textbf{Description}} 
\\ \hline
    transfer & users transfer \textit{STEEM}/\textit{SBD} to other users \\
    transfer\_to\_vesting & users transfer \textit{STEEM}/\textit{SBD} to \textit{VESTS} \\
    delegate\_vesting\_shares & users delegate \textit{VESTS} to other users \\ 
    withdraw\_vesting & users transfer \textit{VESTS} to \textit{STEEM} \\ 
    \hline
\end{tabular}
\end{center}
\caption{Summary of operations}      
\vspace{-9mm}
\label{t1}
\end{table}

\begin{figure*}
\minipage{0.32\textwidth}
  \includegraphics[width=\columnwidth]{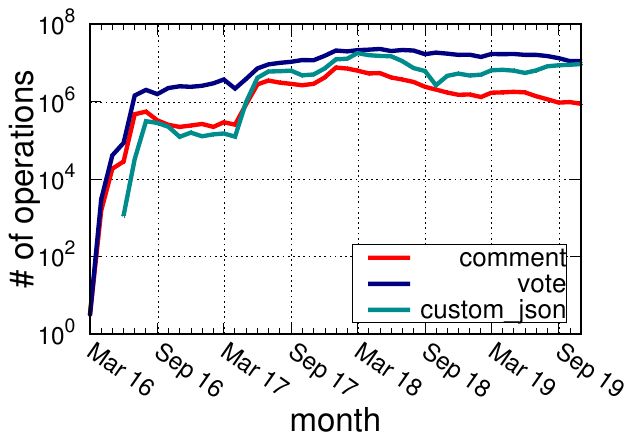}
  \vspace{-8mm}
  \caption {\small New social-network operations per month (2016/03 to 2019/11)}
  \vspace{-2mm}
  \label{sec4_1}
\endminipage\hfill
\minipage{0.32\textwidth}%
  \includegraphics[width=\columnwidth]{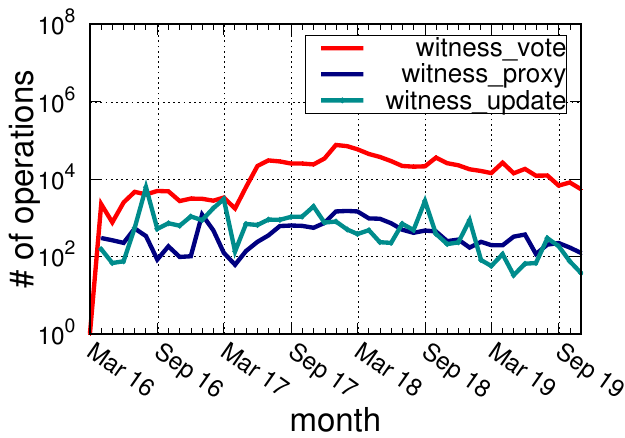}
  \vspace{-8mm}
  \caption {\small New witness-election operations per month (2016/03 to 2019/11)}
  \vspace{-2mm}
  \label{sec4_2}
\endminipage\hfill
\minipage{0.32\textwidth}
  \includegraphics[width=\columnwidth]{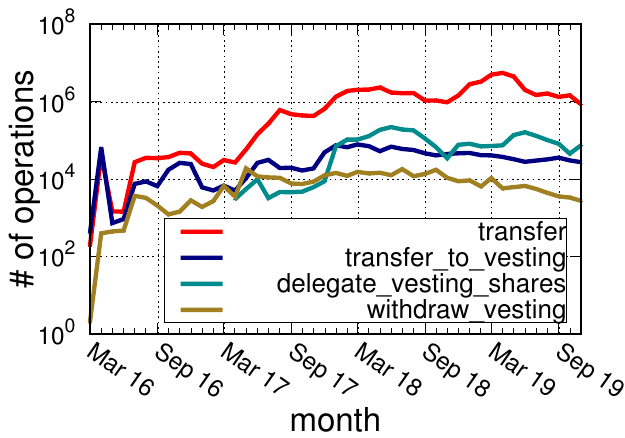}
  \vspace{-8mm}
  \caption {\small New value-transfer operations per month (2016/03 to 2019/11)}
  \vspace{-2mm}
  \label{sec4_3}
\endminipage
\end{figure*}

\section{SteemOps}
In this section, we present \textit{SteemOps}, a new dataset that organizes the key Operations in Steemit. The dataset is available at: \\ 
\centerline{\url{https://github.com/archerlclclc/SteemOps}}

\subsection{Data Extraction}
The Steem-blockchain offers an Interactive Application Programming Interface (API) for developers and researchers to collect and parse the blockchain data~\cite{SteemAPI}.
We collect blockchain data from block 1 (created at 2016/03/24 16:05:00) to block 38,641,150 (created at 2019/12/01 00:00:00). 
In the data collected, we recognized ten key types of operations that are most relevant and useful to research in social networking and blockchain and we classify them into three groups in Table~\ref{t1}.

\subsection{Organization of SteemOps}
\textit{SteemOps} organizes 904,388,432 operations into three sub-datasets corresponding to the three groups of operations in Table~\ref{t1}:
1) social-network operation dataset (SOD);
2) witness-election operation dataset (WOD);
3) value-transfer operation dataset (VOD).
Next, we present our preliminary analysis and describe the dataset in detail.

\subsection{Preliminary analysis}
In Figure~\ref{sec4_1}, Figure~\ref{sec4_2} and Figure~\ref{sec4_3}, we plot the number of social-network operations, witness-election operations and value-transfer operations performed in different months, respectively.
Among the three groups of operations, the social-network operations show the highest utilization rate, which indicates that users are using more social functions offered by \textit{Steemit} than other functions.
Among the three social-network operations, the \textit{vote} operation is the most frequently used one. 
Among the four value-transfer operations, users perform the \textit{transfer} operation more frequently.
Finally, the number of performed witness-election operations is relatively small compared to the other two groups. 

\subsection{Social-network Operation Dataset (SOD)}

The Social-network Operation Dataset (SOD) consists of 92,123,270 \textit{comment} operations, 508,514,846 \textit{vote} operations and 245,859,678 \textit{custom\_json} operations.


\begin{table}
\small
\begin{center}
\begin{tabular}{|p{2cm}|p{1cm}|p{4.5cm}|}
\hline
{\textbf{Field name}} & {\textbf{Type}} & {\textbf{Description}} 
\\ \hline
    block\_no & Integer & the block recording this operation \\
    \hdashline 
    parent\_author & String & the author that comment is being submitted to
    \\
    \hdashline 
    parent\_permlink & String & specific post that comment is being submitted to
    \\
    \hdashline 
    author & String & author of the post/comment being submitted (account name) \\
    \hdashline 
    permlink & String & unique string identifier for the post, linked to the author of the post \\
    \hline
\end{tabular}
\end{center}
\caption{Schema of operation comment}      
\vspace{-8mm}
\label{t4_1}
\end{table}

\subsubsection{\textbf{\textit{comment}}}

This operation in SOD consists of the five fields in Table~\ref{t4_1}. Specifically, when both \textit{parent\_author} and  \textit{parent\_permlink} are empty, the operation indicates a new post. In contrast, when both the two fields are not empty, the operation represents a comment to a post/comment.

\begin{table}
\small
\begin{center}
\begin{tabular}{|p{2cm}|p{1cm}|p{4.5cm}|}
\hline
{\textbf{Field name}} & {\textbf{Type}} & {\textbf{Description}} 
\\ \hline
    block\_no & Integer & the block recording this operation \\
    \hdashline 
    voter & String & voter's account name \\
    \hdashline 
    author & String & author's account name \\
    \hdashline 
    permlink & String & unique string identifier for the post \\
    \hdashline 
    weight & Integer & weight of vote \\
    \hline
\end{tabular}
\end{center}
\caption{Schema of operation vote}      
\vspace{-8mm}
\label{t4_2}
\end{table}

\subsubsection{\textbf{\textit{vote}}}
This operation in SOD includes the five fields in Table~\ref{t4_2}.
It describes that a user has cast a vote with a certain weight on a post/comment.
A user may set voting weight $vw$ to any value between 0\% and 100\%.
\textit{Steemit} leverages voting power $vp$ to restrict the number of weighted votes cast by users per day.
Initially, each user has $vp=100\%$.
Then, if a user keeps voting, his/her $vp$ will keep dropping. 
Each day, $vp$ recovers 20\%.

\begin{table}
\small
\begin{center}
\begin{tabular}{|p{2cm}|p{1cm}|p{4.5cm}|}
\hline
{\textbf{Field name}} & {\textbf{Type}} & {\textbf{Description}} 
\\ \hline
    block\_no & Integer & the block recording this operation \\
    \hdashline 
    required\_posting\_
    auths & String & account name with posting authority \\
    \hdashline 
    id & String & id string with 32 characters at most \\
    \hdashline 
    json & String & the actual payload of the operation, containing a JSON Array \\
    \hline
\end{tabular}
\end{center}
\caption{Schema of operation custom\_json}      
\vspace{-8mm}
\label{t4_3}
\end{table}

\begin{table}
\small
\begin{center}
\begin{tabular}{|p{2cm}|p{1cm}|p{4.5cm}|}
\hline
{\textbf{Field name}} & {\textbf{Type}} & {\textbf{Description}} 
\\ \hline
    block\_no & Integer & the block recording this operation \\
    \hdashline
    owner & String & users who wish to become a witness \\
    \hline
\end{tabular}
\end{center}
\caption{Schema of operation witness\_update}      
\vspace{-7mm}
\label{t4_4}
\end{table}

\subsubsection{\textbf{\textit{custom\_json}}}

This operation in SOD contains the four fields in Table~\ref{t4_3}.
It provides a generic way to post any type of JSON data into the blockchain, such as following, reblog and mute.
When it is used for following, its payload in the json field includes information about both follower's and followee's account names.

\subsection{Witness-election Operation Dataset (WOD)}

The Witness-election Operation Dataset (WOD) consists of 852,896 \textit{witness\_update} operations, 19,555 \textit{witness\_vote} operations and 32,935 \textit{witness\_proxy} operations.
These three types of operations are used in the process of stake-weighted witness election.


\subsubsection{\textbf{\textit{witness\_update}}}
This operation in WOD has the two fields in Table~\ref{t4_4}.
Any user in \textit{Steemit} can run a server, install the Steem-blockchain and synchronize the blockchain data to the latest block.
Then, by sending a \textit{witness\_update} operation to the network, the user can become a witness and have a chance to operate the website and earn producer rewards if he or she can gather enough support from the electors to join the 21-member witness group.


\subsubsection{\textbf{\textit{witness\_vote}}}
This operation in WOD consists of the four fields in Table~\ref{t4_5}.
As a commonly used way to vote for witnesses, a user may choose to perform the \textit{witness\_vote} operation to directly vote for at most 30 witnesses.
It is worth noting that an empty \textit{approve} field means that the user is revoking the vote to the witness.


\subsubsection{\textbf{\textit{witness\_proxy}}}
This operation in WOD includes the three fields in Table~\ref{t4_6}.
As another option to participants in the stake-weighted witness election, a user may choose to perform a \textit{witness\_proxy} operation to set another user as an election proxy.
The weight of a vote is the sum of the voter's own stake and the stake owned by other users who have set the voter as proxy.

\subsection{Value-transfer Operation Dataset (VOD)}

The Value-transfer Operation Dataset (VOD) consists of 52,611,143 \textit{transfer} operations, 1,463,103 \textit{transfer\_to\_vesting} operations, 2,563,749 \textit{delegate\_vesting\_shares} operations and 347,257 \textit{withdraw\_vesting} operations.

\begin{table}
\small
\begin{center}
\begin{tabular}{|p{2cm}|p{1cm}|p{4.5cm}|}
\hline
{\textbf{Field name}} & {\textbf{Type}} & {\textbf{Description}} 
\\ \hline
    block\_no & Integer & the block recording this operation \\
    \hdashline
    account & String & voter's account name \\
    \hdashline 
    witness & String & witness's account name \\
    \hdashline 
    approve & String & arppove a new vote or revoke an old vote \\
    \hline
\end{tabular}
\end{center}
\caption{Schema of operation witness\_vote}      
\vspace{-7mm}
\label{t4_5}
\end{table}

\begin{table}
\small
\begin{center}
\begin{tabular}{|p{2cm}|p{1cm}|p{4.5cm}|}
\hline
{\textbf{Field name}} & {\textbf{Type}} & {\textbf{Description}} 
\\ \hline
    block\_no & Integer & the block recording this operation \\
    \hdashline
    account & String & user's account name \\
    \hdashline 
    proxy & String & proxy's account name \\
    \hline
\end{tabular}
\end{center}
\caption{Schema of operation witness\_proxy}      
\vspace{-7mm}
\label{t4_6}
\end{table}

\begin{table}
\small
\begin{center}
\begin{tabular}{|p{2cm}|p{1cm}|p{4.5cm}|}
\hline
{\textbf{Field name}} & {\textbf{Type}} & {\textbf{Description}} 
\\ \hline
    block\_no & Integer & the block recording this operation \\
    \hdashline
    from & String & sender's account name \\
    \hdashline 
    to & String & recipient's account name \\
    \hdashline
    amount & String & the amount of transferred asset \\
    \hdashline 
    memo & String & a memo string with 2048 bytes at most \\
    \hline
\end{tabular}
\end{center}
\caption{Schema of operation transfer}      
\vspace{-8mm}
\label{t4_7}
\end{table}

\subsubsection{\textbf{\textit{transfer}}}
This operation in VOD includes the five fields in Table~\ref{t4_7}.
It is used for transferring asset from one account to another. 
With this operation, a user can transfer either STEEM or SBD, but transferring of Steem Power (VESTS) is not allowed.
A sender can leave a short message in the \textit{memo} field but one needs to pay attention that the memo is plain-text.

\subsubsection{\textbf{\textit{transfer\_to\_vesting}}}
This operation in VOD consists of the four fields in Table~\ref{t4_8}.
It is used for converting STEEM into VESTS at the current exchange rate.
A user can either leave the \textit{to} field empty to receive the VESTS, or set the \textit{to} field to the recipient's account name to transfer VESTS to another account.
The latter usage of \textit{transfer\_to\_vesting} allows faucets to pre-fund new accounts with VESTS.

\subsubsection{\textbf{\textit{delegate\_vesting\_shares}}}
This operation in VOD is formed by the four fields in Table~\ref{t4_9}.
It is used for delegating VESTS from one account (i.e., delegator) to the other (i.e., delegatee). 
It is worth emphasizing that the delegated VESTS are still possessed by the delegator, who can increase or decrease the amount of delegated VESTS at any time and even completely remove the delegation by setting the \textit{vesting\_shares} field to zero.
Upon receiving delegated VESTS, the delegatee could leverage the amount of delegated VESTS to increase the power of votes to contents such as blogs and comments.




\begin{table}
\small
\begin{center}
\begin{tabular}{|p{2cm}|p{1cm}|p{4.5cm}|}
\hline
{\textbf{Field name}} & {\textbf{Type}} & {\textbf{Description}} 
\\ \hline
    block\_no & Integer & the block recording this operation \\
    \hdashline
    from & String & sender's account name \\
    \hdashline 
    to & String & recipient's account name \\
    \hdashline
    amount & String & the amount of vested asset \\
    \hline
\end{tabular}
\end{center}
\caption{Schema of operation transfer\_to\_vesting}      
\vspace{-8mm}
\label{t4_8}
\end{table}

\begin{table}
\small
\begin{center}
\begin{tabular}{|p{2cm}|p{1cm}|p{4.5cm}|}
\hline
{\textbf{Field name}} & {\textbf{Type}} & {\textbf{Description}} 
\\ \hline
    block\_no & Integer & the block recording this operation \\
    \hdashline
    delegator & String & delegator's account name \\
    \hdashline 
    delegatee & String & delegatee's account name \\
    \hdashline
    vesting\_shares & String & the amount of delegated VESTS\\
    \hline
\end{tabular}
\end{center}
\caption{Schema of operation delegate\_vesting\_shares}      
\vspace{-8mm}
\label{t4_9}
\end{table}

\begin{table}
\small
\begin{center}
\begin{tabular}{|p{2cm}|p{1cm}|p{4.5cm}|}
\hline
{\textbf{Field name}} & {\textbf{Type}} & {\textbf{Description}} 
\\ \hline
    block\_no & Integer & the block recording this operation \\
    \hdashline
    account & String & withdrawer's account name \\
    \hdashline
    vesting\_shares & String & the amount of VESTS to withdraw\\
    \hline
\end{tabular}
\end{center}
\caption{Schema of operation withdraw\_vesting}      
\vspace{-8mm}
\label{t4_10}
\end{table}

\subsubsection{\textbf{\textit{withdraw\_vesting}}}
This operation in VOD consists of the three fields in Table~\ref{t4_10}.
It is used by users to withdraw their VESTS at any time.
It is worth noting that the VESTS have to be withdrawn in the next thirteen subsequent weeks.
For example, in day 1, Alice may invest 13 \textit{STEEM} to \textit{Steemit} that makes her vote obtain a weight of 13 \textit{SP} (about 26000 \textit{VESTS}). Later, in day 8, Alice may decide to withdraw her 13 invested \textit{STEEM}. Here, instead of seeing her 13 \textit{STEEM} in wallet immediately, her \textit{STEEM} balance will increase by 1 \textit{STEEM} each week from day 8 and during that period, her \textit{SP} will decrease by 1 \textit{SP} every week.




\section{Applications of SteemOps and related work}
The unique aspect of \textit{SteemOps}, namely the deep integration of the underlying social network and blockchain, can support a diverse set of potential applications for researchers in both the communities and even in other domains such as economics~\cite{jeong2020centralized,kim2019sustainable}.
In this section, we present some notable research opportunities based on \textit{SteemOps} and their related work.

\subsection{Blockchain System Analysis}
We first discuss three key research opportunities based on \textit{SteemOps} on blockchain system analysis.

\subsubsection{\textbf{Decentralization analysis}}
Decentralization is a key indicator for the evaluation of public blockchains. 
Most existing works on decentralization in blockchains have focused on Bitcoin~\cite{beikverdi2015trend,eyal2015miner,eyal2014majority,chao2021measuring}. 
These works pointed out that Bitcoin shows a trend towards centralization because of the emergence of mining pools.
In~\cite{eyal2014majority}, the authors proposed the notion of selfish mining, which reduces the bar for performing 51\% attack to possessing over 33\% of computational power in Bitcoin.
Later, authors in~\cite{eyal2015miner} analyzed the mining competitions among mining pools in Bitcoin from the perspective of game theory and proposed that a rational mining pool may get incentivized to launch a block withholding attack to another mining pool.
Besides Bitcoin, recent work has analyzed the degree of decentralization in Steem~\cite{li2019incentivized}. The work analyzed the process of witness election in Steem from the perspective of network analysis and concluded that the Steem network was showing a relatively low level of decentralization.
Recently, there have been a few studies on comparing the level of decentralization between different blockchains, such as Bitcoin/Ethereum~\cite{gencer2018decentralization} and Bitoin/Steem~\cite{li2020comparison,kwon2019impossibility}.
Specifically, the degree of decentralization in Steem was computed among witnesses in~\cite{kwon2019impossibility}, which may fail to reflect the actual degree of decentralization in a DPoS blockchain. 
Later, Li \textit{et al.}~\cite{li2020comparison} quantified the degree of decentralization in Steem from the perspective of stakeholders after analysis and measurements of the witness election.
With the rich operations offered by \textit{SteemOps}, the degree of decentralization in Steem blockchain could be further analyzed from more perspectives such as among voters, authors and proxies.

\subsubsection{\textbf{Cryptocurrency transfer analysis}}
In recent years, the cryptocurrency transferring networks have become the main resources for supporting a number of empirical studies.
Yousaf \textit{et al.}~\cite{yousaf2019tracing} used data from 
ShapeShift platform and eight different blockchains to explore whether or not money can be traced as it moves across ledgers, and their results identified various patterns of cross-currency trades.
Lee \textit{et al.}~\cite{lee2019cybercriminal} extracted cryptocurrency information related to Dark Web and analyzed their usage characteristics on the Dark Web.
Chen \textit{et al.}~\cite{chen2019market} analyzed the leaked transaction history of Mt. Gox Bitcoin exchange and concluded that there was serious market manipulation in Mt. Gox exchange and
the cryptocurrency market must strengthen the supervision. 
Chen \textit{et al.}~\cite{chen2020traveling} conducted a systematic investigation on the whole Ethereum ERC20 token ecosystem to characterize the token creator, holder, and transfer activity. 
\textit{SteemOps} offers rich value-transfer operations, including transferring of STEEM, transferring to VESTS and delegating VESTS and thus facilitates various angles of analysis.

\subsubsection{\textbf{Performance benchmark}}
Many recent new blockchain systems such as Omniledger~\cite{kokoris2018omniledger} and Monoxide~\cite{wang2019monoxide} aim at improving the performance of blockchains, thus requiring real transaction data collected from existing blockchain systems to evaluate their solutions.
For instance, Monoxide leveraged historical transaction data in Ethereum in its evaluation.
To support such requirements, performance benchmarks such as Blockbench~\cite{dinh2017blockbench} have been proposed, but most of the existing benchmarks create workloads by simulating user behaviors, which may not well match with the real data and may decrease the accuracy of the evaluation results.
In contrast, \textit{SteemOps} provides a substantial number of well-processed operations that cover different aspects of a blockchain system including DPoS consensus protocol and cryptocurrency ecosystem.
It is worth noting that the DPoS-powered \textit{Steemit} social media platform leverages a small set of witnesses that are periodically elected by the entire stakeholder community to boost the transaction throughput and therefore, it can serve as a state-of-the-art workload for comparison in evaluation.

\subsection{Social Network Analysis}
Next, we identify some key research opportunities based on \textit{SteemOps} on social network analysis.

\subsubsection{\textbf{Community and user behavior analysis}}

In the past few years, due to their rapid growth and consistent popularity, social media platforms have received significant attention from researchers. 
A great number of research papers have analyzed the community and user behavior in many popular social media platforms.
Tan \textit{et al.}~\cite{tan2015all} investigated user behavior in \textit{Reddit} and found that users continually post in new communities.
Singer \textit{et al.}~\cite{singer2016evidence} observed a general quality drop of comments made by users during activity sessions.
Hessel \textit{et al.}~\cite{hessel2016science} investigated the interactions between highly related communities and found that users engaged in a newer community tend to be more active in their original community.
In~\cite{glenski2017consumers}, the authors studied the browsing and voting behavior of \textit{Reddit} users and found that most users do not read the article that they vote on.
Wang \textit{et al.}~\cite{wang2013wisdom} analyzed the \textit{Quora} platform and found that the quality of Quora’s knowledge base is mainly contributed by its user heterogeneity and question graphs.
Anderson \textit{et al.}~\cite{anderson2012discovering} investigated the \textit{Stack Overflow} platform and observed significant assortativity in the reputations of co-answerers, relationships between reputation and answer speed.
With \textit{SteemOps}, especially its Social-network Operation Dataset (SOD), researchers without any blockchain system background can easily leverage the well-processed operations to analyze community and user behavior in \textit{Steemit} and compare their results with that in other social media platforms. 
It would be also very interesting to understand the impacts that blockchains may bring on users' social behavior.

\subsubsection{\textbf{Curation mechanism}}
In \textit{Steemit}, users create content as posts that get curated based on votes from other users. 
The platform periodically issues cryptocurrency as rewards to creators and curators of popular posts.
Thelwall \textit{et al.}~\cite{thelwall2017can} analyzed the first posts made by 925,092 \textit{Steemit} users to understand the factors that may drive the post authors to earn higher rewards.
Their results suggest that new users of \textit{Steemit} start from a friendly introduction about themselves rather than immediately providing useful content.
In a very recent work, Kiayias \textit{et al.}~\cite{kiayias2018puff} studied the decentralized content curation mechanism from a computational perspective. They defined an abstract model of a post-voting system, along with a particularization inspired by \textit{Steemit}.
Through simulation of voting procedure under various conditions, their work identified the conditions under which \textit{Steemit} can successfully curate arbitrary lists of posts and also revealed the fact that selfish participant behavior may hurt curation quality.
Compared with existing works, the rich historical data collected in \textit{SteemOps} would offer researchers a deep and insightful view on the detailed stake-weighted voting procedure that determines the amount of curation authors earn. 

\subsubsection{\textbf{Bot detection}}
The rise of social bots and the harm caused by them to the online ecosystems has been widely recognized~\cite{ferrara2016rise}.
In \textit{Steemit}, although its reward system is originally driven by the desire to incentivize users to contribute high-quality content, the analysis of the underlying cryptocurrency transfer network on the blockchain in a recent work~\cite{li2019incentivized} reveals that more than 16\% transfers of cryptocurrency in \textit{Steemit} are sent to curators suspected to be bots. The study also finds the existence of an underlying supply network for the bots suggesting a significant misuse of the current reward system in \textit{Steemit}.
\textit{SteemOps} offers rich data on detecting bots, such as memo information carried by \textit{transfer} operations and correlations between accounts revealed by \textit{delegate\_vesting\_shares} operations.
The transparency of its social network and cryptocurrency network could facilitate a better understanding of bots in social media platforms.

\section{Conclusion}
This paper presents \textit{SteemOps}, a new dataset that organizes over 900 million operations from \textit{Steemit} into three sub-datasets:
1) social-network operation dataset (SOD);
2) witness-election operation dataset (WOD);
3) value-transfer operation dataset (VOD).
In \textit{SteemOps}, we collect over 38 million blocks generated during 45 months from 2016/03 to 2019/11 and extract ten key types of operations performed by \textit{Steemit} users from blocks.
We describe the dataset schema information and its usage in detail and outline various potential research directions based on \textit{SteemOps}.
We believe that \textit{SteemOps} can facilitate impactful future studies
and can support a diverse set of potential applications for researchers in both the social networking and blockchain research communities.

\section*{Acknowledgement}
Chao Li is partially supported by Fundamental Research Funds for the Central Universities (No. 2019RC038).\

\renewcommand\refname{Reference}

\bibliographystyle{plain}
\urlstyle{same}

\bibliography{main.bib}

\end{document}